\documentclass[pdflatex,sn-chicago]{sn-jnl}
\usepackage{graphicx}%
\usepackage{multirow}%
\usepackage{amsmath,amssymb,amsfonts}%
\usepackage{amsthm}%
\usepackage{natbib}
\usepackage{mathrsfs}%
\usepackage[title]{appendix}%
\usepackage{xcolor}%
\usepackage{textcomp}%
\usepackage{manyfoot}%
\usepackage{booktabs}%
\usepackage{algorithm}%
\usepackage{algorithmicx}%
\usepackage{algpseudocode}%
\usepackage{listings}%
\usepackage{bm}
\usepackage{anyfontsize}

\begin{document}

\title[]{ Geographical inequalities in mortality by age and gender in Italy, 2002-2019: insights from a spatial extension of the Lee-Carter model.}


\author[1]{\fnm{Francesca} \sur{Fiori}}

\author[2]{\fnm{Andrea} \sur{Riebler}}

\author[2]{\fnm{Sara} \sur{Martino}}

\affil[1]{\orgname{University of Strathclyde},  \country{UK}}

\affil[2]{\orgname{Norwegian University of Science and Technology}, \country{Norway}}


\abstract{
Italy reports some of the lowest levels of mortality in the developed world. Recent evidence, however, suggests that even in low-mortality countries improvements may be slowing and regional inequalities widening.
This study contributes new empirical evidence to the debate by analysing mortality data by single year of age for males and females across 107 provinces in Italy from 2002 to 2019. We extend the widely used Lee–Carter model to include spatially varying age-specific effects, and further specify it to capture space–age–time interactions. The model is estimated in a Bayesian framework using the \texttt{inlabru} package, which builds on INLA (Integrated Nested Laplace Approximation) for non-linear models and facilitates the use of smoothing priors. This approach borrows strength across provinces and years, mitigating random fluctuations in small-area death counts.
Results demonstrate the value of such a granular approach, highlighting the existence of an uneven geography of mortality despite overall national improvements. Mortality disadvantage is concentrated in parts of the Centre–South and North–West, while the Centre–North and North–East fare relatively better. These geographical differences have widened since 2010, with clear age- and gender-specific patterns, being more pronounced at younger adult ages for men and at older adult ages for women. 
Future work may involve refining the analysis to mortality by cause of death or socioeconomic status, informing more targeted public health policies to address mortality disparities across Italy’s provinces.
}




\maketitle

\section{Introduction}\label{sec1}

Over the past century, high-income countries have seen remarkable improvements in survival, driven by advancements in living conditions, medical innovations, healthcare technologies, behavioural changes, and public health policies \citep{omran71, VallinMesle}. However, these gains have not been distributed evenly between and within countries \citep{VallinMesle, Sauerberg2024, Hrzic2020} or across different social groups \citep[e.g.,][]{MACKENBACH2012, MACKENBACH2017}. Moreover, survival improvements in the last 15 years have slowed down in most countries, and even reversed in some of the worst performers \citep{KabirObrien, Dowd2024}, raising the question of where, and for whom, progress has slowed or come to a standstill.

A growing body of recent research has paid attention to differences in life expectancy within individual European countries, revealing the existence of significant sub-national variation (e.g., in Spain  \citep{Bramajo23}; in Finland and Sweden \citep{wilson2020}; in Greece \citep{Tsimbos2014}; in Belgium \citep{OTAVOVA2023100587}; in Germany  \citep{vanRaalte2020}—among others). Systematic comparison of literature \citep{Hrzic2020}, and of European-wide data \citep{Sauerberg2024}, indicate that the slowdown has indeed been accompanied by an increasing dispersion across geographical units,
even within countries with historically low mortality rates. 

Italy presents a compelling case study in mortality trends. While the country reports some of the lowest mortality levels in Europe, with a life expectancy at birth of 83.4 years in 2023—surpassing the EU27 average of 81.4 \citep{eurostat25}—this remarkable overall achievement conceals the existence of geographical inequalities. According to the most recent estimates, for example, nearly four years differences in life expectancy at birth were observed between provinces in the South (e.g., 80.9 in Caserta – Campania) and in the Centre-North (e.g., 84.7 in Firenze – Toscana) \citep{website:istat24}. These differences in mortality likely reflect the existence of well-documented geographical disparities in wealth, educational levels and employment opportunities \citep{Danile_malamina}; in the concentration of economic activities such as heavy manufacturing industry, with associated environmental risks \citep{De_dominicis}; in the availability and quality of healthcare services associated with a decentralised health system \citep{balia2018, benassi2025}.

Moreover, despite its comparatively low mortality levels, the country has experienced some slowdown in the pace of mortality decline since 2010. This deceleration has been linked to the diminishing rate of progress in reducing cardiovascular disease-related deaths \citep{oecd2020}, the “double jeopardy” \citep{Abramsetal2023} of slower improvements at both younger (25-40) and older (55-75) ages \citep{Dowd2024}, and smaller-than-expected gains in longevity for women \citep{Djeundje22}. Changes in mortality patterns by age, gender and underlying cause, may also find expression geographically, with some areas and population groups experiencing slower progress than others.

This study intends to make an empirical and methodological contribution to understandings of inequalities in contemporary mortality trends in high-income countries. First, it uses mortality data for 107 Italian provinces to study geographical variations in mortality trends beyond population aggregates, studying age-specific mortality rates by gender. Second, it demonstrates the opportunities offered by Bayesian spatial approaches \citep{goes-2024} and by recent computational advancements \citep{rue2009approximate} to overcome the challenges of producing estimates at a fine geographical detail broken down by one or more characteristics (such as age, gender, time).
Specifically, it estimates a spatial extension of the traditional Lee-Carter model \citep{lee-carter92, BASELLINI2023}  to address the following research questions:
\begin{enumerate}
    \item How has mortality by age and gender evolved over the past two decades?
    \item Are there geographical inequalities in mortality rates by age and gender?
    \item And have geographical inequalities widened during the slowdown of survival improvement of the 2010s?
\end{enumerate}

\section{The Changing Geography of Mortality in Italy}\label{sec2}

Demographic research has extensively documented geographical differences in mortality across Italy during the second half of the 20th century \citep[e.g.,][]{capocaccia1990, CaselliEgidi1979, CaselliEgidi1981, Caselli1983,CaselliReale1999, caselli2003a}. While highlighting general improvements in survival, this body of research reveals a distinct gendered geography of mortality. Among men, high mortality was initially concentrated in the North—largely due to environmental risks associated with rapid industrialization—but later spread to parts of the Centre-South, particularly Campania. In contrast, the South and the Islands remained the regions with the greatest advantage. For women, mortality, especially among the elderly, was highest in the Apennine region, Sicily, and parts of the North, while the Centre-North was the least affected. Since the 1970s, the decline in mortality has reshaped these patterns. Although the North-South divide persisted for men, high-mortality zones in the North contracted, while Campania and Sicily became new areas of concern. For women, mortality patterns diverged from those of men, with the highest rates now concentrated in Campania and Sicily, particularly among the elderly \citep{caselli2003a}. Cause-specific mortality trends further qualify these gendered patterns. In the wealthier North, men exhibited higher rates of cancer and ischemic diseases, whereas in the less affluent South, mortality was elevated for conditions that could be mitigated through better healthcare. Among women, the South recorded the highest mortality from circulatory diseases and diabetes, while cancer and ischemic heart disease played a lesser role \citep[e.g., ][]{ CaselliEgidi1981, caselli2003a}.

Collectively these studies demonstrate that geographical analyses can yield valuable insights, particularly for informing public policy. When combined with analyses of temporal changes, and focused on specific diseases or population age groups, they suggest plausible influence of underlying social, cultural, and environmental factors in different areas. Drawing on observed trends, some of these studies \citep{caselli2003a} concluded by suggesting that the following century would have ushered a new geographical landscape, particularly for adult and elderly male mortality, reinforcing the need for continued research into spatial variation.

Surprisingly then, comparatively little scholarly attention to geographical variations in mortality across Italy followed up in the early 21st century. A few notable exceptions include recent work by \citet{carboni2024} and \citet{camarda25}, which analysed spatial convergence in life expectancy using data up to 2019. Their studies suggest a reversal of previous convergence trends, with mortality disparities widening in recent years, particularly at younger adult ages. Although the Northern and Central regions—particularly the North East—have experienced greater improvements in life expectancy, the South and the Islands have lagged behind. \citet{carboni2024} attribute this shift to healthcare decentralization and financial constraints, especially following the 2008 economic crisis, which disproportionately affected poorer Southern regions.
Similar findings emerge from \citet{Sauerberg2024}, who examined provincial data from 16 European countries, including Italy. Their study confirms that despite overall improvements, sub-regional disparities have increased in recent years. Notably, mortality rates for men in previously disadvantaged areas of Northern Italy have aligned with national averages, while Southern regions continue to experience higher mortality rates for both genders \citep{Culotta2021}. Similarly, \citet{santossanchez2020}, using municipal data from 2012 to 2016, identified mortality hotspots in Southern Italy linked to high unemployment levels.

While these studies provide important insights on features and changes of the recent geography of mortality in Italy, they are limited in the extent to which they isolate the contribution of specific aspects to overall patterns. A recent reflection paper on the present and future of mortality studies \citep{Dowd2024} emphasized the important contribution of age- and cause-specific mortality to understand current and future trends, and the extent to which they reflect biological limits or social, environmental and structural factors that are amenable to change.

Indeed, studies that break down mortality patterns by age and/or cause can provide valuable insights into the underlying drivers \citep{caselli2003a}. \citet{Barbietal2018} argue that mortality by leading cause of death can offer a more accurate and nuanced measure of regional differences in economic development and social wellbeing. Their analyses demonstrate that the geography of mortality from cardiovascular diseases and diabetes reflects the well-known North-South divide, with Southern regions faring worse for their combination of lower levels of socio-economic prosperity and poorer healthcare provision \citep{benassi2025}. A completely different picture emerges by mapping mortality rates from cancer, which instead highlights the disadvantage, particularly for men, of the more industrialised and economically developed Northern regions. \cite{divino_egidi} illustrated similar patterning with a finer level of detail, breaking down mortality rates at the provincial level by gender, leading cause, and large age-class. Using a Bayesian spatial approach, which borrows information from neighbouring provinces, they were able to produce reliable and accurate estimates of mortality rates which would have otherwise been subject to random fluctuations, implicit when studying small population and relatively rare events.


\section{The Data} \label{sec:data}

We analyse  mortality data by single year of age (0 to 95 years) for males and females in Italy for the period 2002 to 2019. Data on overall death and populations counts come from populations registers, and refer to $107$ Italian Provinces (NUTS3), a geographical and administrative division of the country, with population sizes ranging from over 4 million to less than 80 thousand residents according to the last census.
As it is often the case, administrative boundaries vary over time, and some adjustments were necessary to obtain a consistent classification of counts over time. In particular, the series on death counts was reconstructed by ISTAT (Italian National Statistics Office) for the purpose of this study, so that, for both series,  
the geography is consistent over time and counts refer to the most recent territorial classification.
As data come from Population registers, they can be considered precise and of good quality. However, as they refer to territorial units with varying population sizes, they are subject to some degree of random fluctuations, hence the value of a Bayesian framework. 

\section{A spatial extended Lee-Carter model}


Estimating mortality rates below national levels presents significant challenges, particularly when estimates are broken down by one or more characteristics (such as time, age, gender or causes of death) and/or refer to a very fine geographical detail. The resulting spatial distribution of mortality risk may be heavily influenced by random variations present in the observed data. This randomness can conceal true patterns and lead to misleading conclusions, especially when events are rare. Two different approaches have been used in the literature to overcome this problem, and thus reduce variability in the estimates: the first involves pooling together information, e.g., by averaging across multiple years and/or reducing the level of detail of the estimates; the second relies on borrowing ‘strength’ from related observations (e.g., data points that are close in time and/or space).

The method we propose in this paper falls within the second group. We apply an extended version of the classical Lee-Carter model \citep{lee-carter92,BASELLINI2023}, a functional model traditionally used to estimate and forecast age patterns of mortality, to the study of the geographical and temporal variation of age- and gender-specific mortality patterns in Italy. Inference is performed within a Bayesian framework using smoothing priors. 
We use the newly developed R-package \texttt{inlabru} \citep{inlabru}, which  extends the popular INLA (Integrated Nested Laplace Approximation) methodology \citep{rue2009approximate} to models with multiplicative effects such those characterizing the Lee-Carter model.  The inlabru approach offers a valuable alternative to Markov Chain Monte Carlo (MCMC) techniques by enabling fast and accurate approximate inference. It provides an intuitive framework for specifying complex models, thereby lowering the barrier for applied researchers to adopt Bayesian inference in practice. 
We demonstrate the value of applying a Bayesian extension of the Lee-Carter Model using mortality data for Italy for the period 2002 to 2019.

We fit a Poisson version of the Lee Carter model \citep{brouhns2002poisson}, with a spatial extension, similar to what proposed in \citet{goes-2024}, but further interacted by age group, to the mortality data introduced in Section \ref{sec:data}, independently for males and females. Let $Y_{xts}$ represent the number of deaths for age $x$, at time $t$ in province $s$ with
    $Y_{xts} \mid \lambda_{xts}\sim\text{Poisson}(E_{xts}\lambda_{xts})$, 
where $E_{xts}$ is the population at risk and $\lambda_{xts}$ the death rate. We model the log rates as:
\begin{equation}\label{eq:model1}
    \log \lambda_{xts} = \alpha_x + \beta_x\kappa_t + \omega_{sg_x} + z_{xts}
\end{equation}
where $\alpha_x$ is a age profile at age $x$, $\kappa_t$ a time effect and $\beta_x$ an age-specific multiplication factor.  Further, $\omega_{sg_x}$ represents an age-specific spatial effect, where $g_x$ indicates the 10 year age group (0--10, 11--20, 21--30\dots) in which age $x$ falls into. Adjustment for overdispersion is incorporated directly in the model by the inclusion of independent Gaussian random variables $z_{xts} \sim \mathcal{N}(0, \sigma_z^2)$. As in \cite{Wisniowski2015} we assume $\alpha_x$ and $\beta_x$ to be mean zero Gaussian random effects with large fixed variance. For the $\kappa_t$'s we chose a random walk of second order whose directed formulation is given as:
\begin{equation} \label{eq:drift}
 \kappa_t = 2 \kappa_{t-1} - \kappa_{t-2} + \epsilon_{t}, \: t=3, \ldots T.
\end{equation}
Here, we assume a uniform prior  for $\kappa_1$ and $\kappa_2$. The error terms $\epsilon_t$, $t=1, \ldots, T$, are independently and identically Gaussian distributed with mean zero and standard deviation parameter $\sigma_\kappa$. Notably, the second order random walks is conceptually similar to a first-order random walk with a time-varying drift component. 
For each age group $g_x$, the $\omega_{sg_x}$´s  follow a BYM2 model \citep{riebler2016intuitive}:
\begin{equation}
\omega_{sg_x}=  \sigma_\omega \left( \sqrt{1 - \phi} \cdot v_{sg_x} + \sqrt{\phi} \cdot u_{sg_x}\right), 
\label{eq:bym2}
\end{equation}
where $v_{sg_x} \sim \mathcal{N}(0, 1)$ and $\mathbf{u}_{g_x}$ follow a scaled Besag model allowing spatial smoothing between provinces \citep{riebler2016intuitive}. In fact, the Besag model assumes that the mean of the province $s$ conditional on all other provinces is given by the average over its neighboring provinces, and the variance is scaled by the number of neighbors. To account for the fact that the graph for Italy is disconnected, we use adjustments as proposed in \cite{FRENISTERRANTINO201825}. The mixing parameter $\phi \in [0, 1]$  measures the proportion
of marginal variance $\sigma_\omega^2$ explained by the spatially structured effect $\bm{u}_{g_x}$, for details on the BYM2 models, we refer to \citet{riebler2016intuitive}.
To ensure identifiability of the model parameters, standard constraints are imposed such that the sum of $\beta_x$ over age is 1, the sum of $\kappa_t$ over time and the sum of $\omega_{sg_x}$ over space is 0 for every age group $g_x$ \citep{lee-carter92}.

In a second model specification we modify model \eqref{eq:model1} to allow the spatial effect $w_{sg_x}$ to depend on time. We use $w_{sg_xp_t}$ in \eqref{eq:bym2}, where $p_t$ indicates one of two time intervals, 2002--2010 or 2011--2019, in which time index $t$ falls into. This enables us to address our third research question of whether the geography of mortality remained constant or changed over time. The cutoff point for the time period responds to both empirical and substantive considerations: first, it divides the overall period in two intervals of equal length; second, 2010 corresponds to the time a slowdown of mortality improvement post economic recession starts to be observed elsewhere in the industrialised world \citep{Dowd2024}.

Performing Bayesian inference we assign a priori distributions to all parameters. We use recently proposed penalized complexity priors for the variance parameters of all random effects \citep{simpson2017penalising} as well as the mixing parameters $\phi$. 
    
\section{Results}

\subsection{Mortality variation by age and gender}


Through the estimation of the Lee-Cartel model on mortality data for the period 2002-2019, we address the first research question, highlighting the existence of gender- and age-specific trends in the decline of mortality over the period considered.

\begin{figure}
    \centering
    \includegraphics[width=\textwidth]{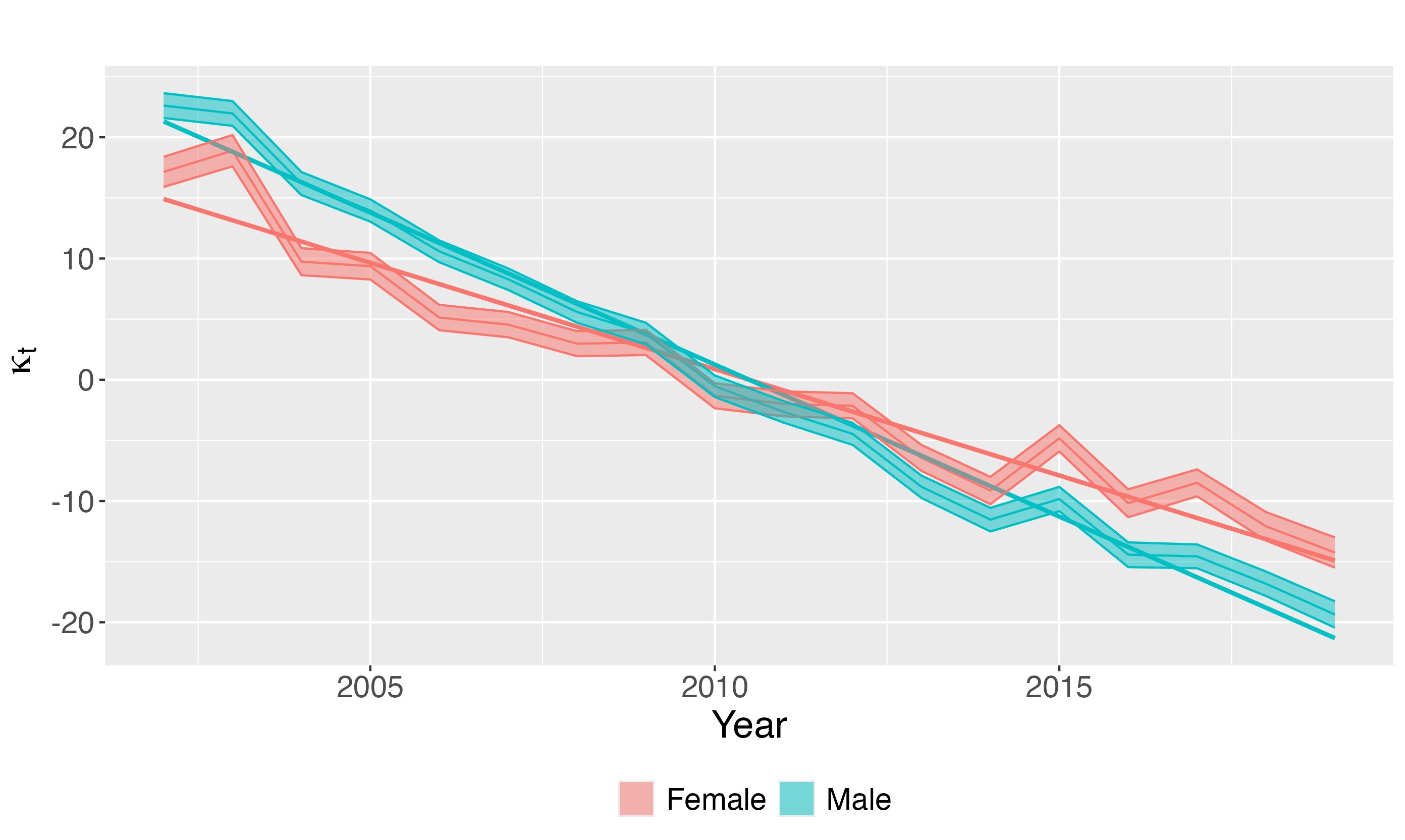}
    \caption{Estimated time trend $\kappa_t$ for males and females.The plot displays the posterior mean (line) together with 95\% credible intervals (shaded area) for both males and females. The two straight lines represent a hypothetical situation of constant rates of mortality decline}
    \label{fig:time_effect}
\end{figure}

Figure \ref{fig:time_effect} shows the estimated $\kappa_t$, a time index of the general level of mortality. Generally, a linear $\kappa_t$  well captures the historical decline in mortality, and corresponds to a decelerating increase in life expectancy at birth. It also indicates that mortality rates have been decreasing exponentially at their own constant rate over the period of observation. 
Between 2002 and 2019, the general level of mortality decreased for both genders, although at a faster pace for men—as indicated by the steeper slope of the blue curve. Moreover, the deviation of the estimated $\kappa_t$ from a perfectly linear fit suggests an accentuation of the slowdown in the decline of mortality, particularly after 2014. This accentuated deceleration seems more pronounced for women.

\begin{figure}
    \centering
    \includegraphics[width=\textwidth]{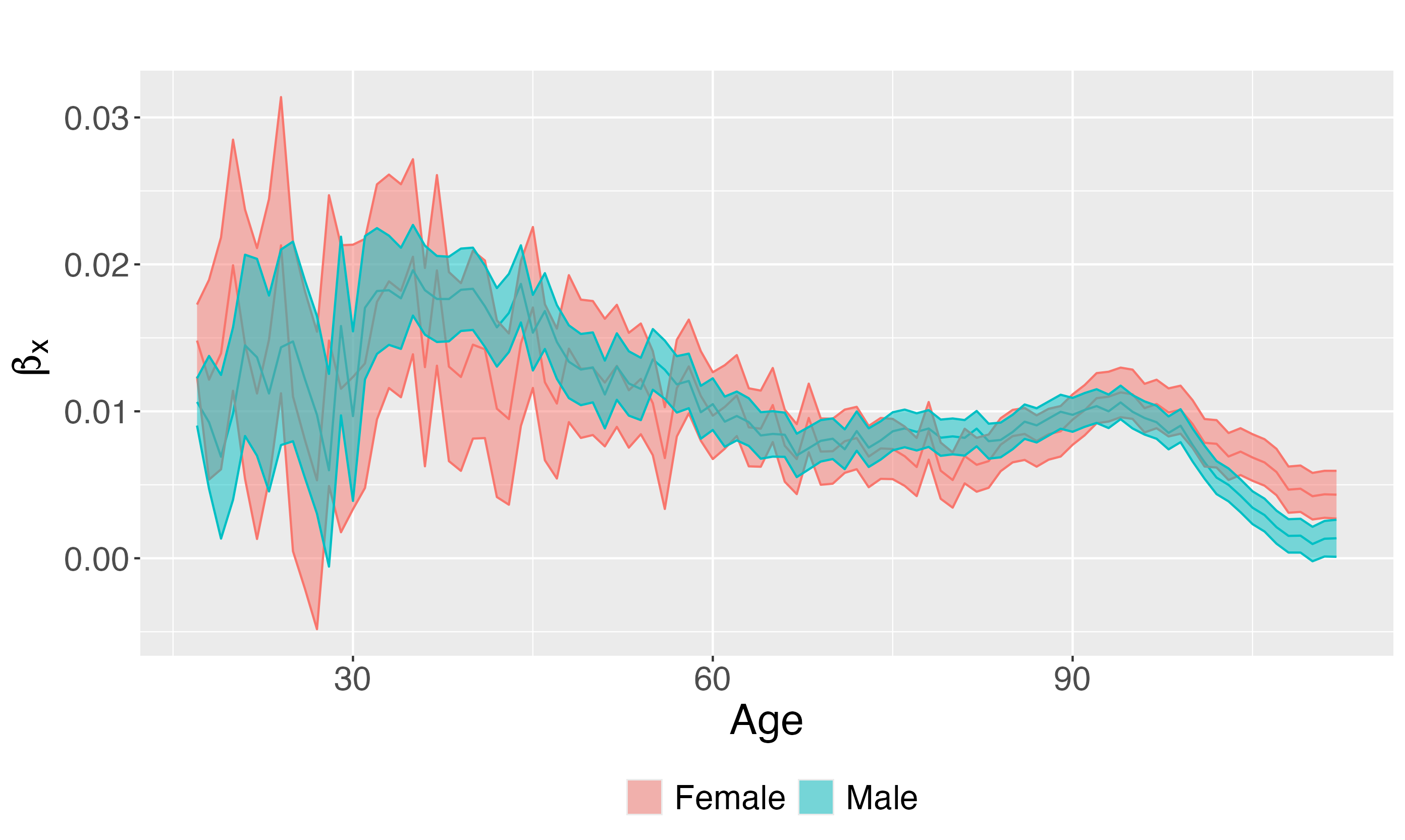}
    \caption{Estimated values for the age-specific multiplicator $\beta_x$: higher values are associated with faster mortality change. The plot displays the posterior mean (line) together with 95\% credible intervals (shaded area) for both males and females.}
    \label{fig:Age_effect}
\end{figure}

Figure \ref{fig:Age_effect} displays the estimated $\beta_x$, a parameter that captures the extent of changes in mortality at age $x$, given the overall temporal trend of overall mortality. Higher values of $\beta_x$ indicate a faster change in mortality. 
Not surprisingly, the decline in mortality rates has been faster at younger ages, although this is also when mortality events tend to be rarer and their estimates subject to greater uncertainty. The pace of mortality decline slows down through adult ages, before rebounding after age 75 (and up to age 95).
The age pattern of changes in mortality is quite similar between the two genders, but there are some differences worth noting. The decline has been faster for men than for women at young adult ages (35 to 45) and between 75 and 85, i.e., at ages in which mortality levels for men are higher than for women. Conversely, after age 90, improvements have been faster for women.

\begin{figure}
    \centering
    \includegraphics[width=\textwidth]{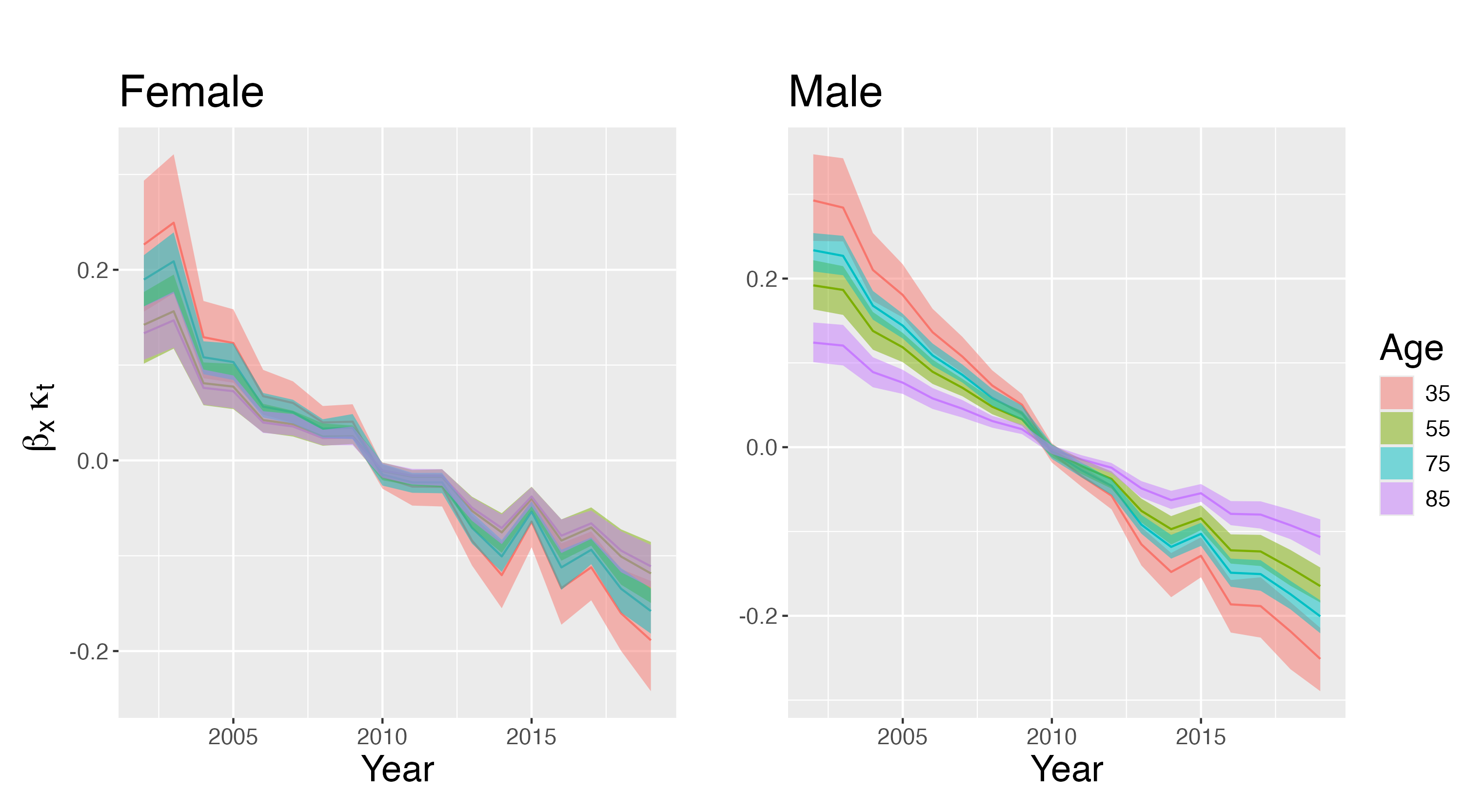}
    \caption{Estimated compound age-specific time effect; $beta_x\kappa_t$. The plot displays the posterior mean (line) together with 95\% credible intervals (shaded area) for both males and females and for selected ages.}
    \label{fig:Time_Age_effect}
\end{figure}

Figure \ref{fig:Time_Age_effect} facilitates the visualization of gender differences in age-specific temporal trends in mortality rates by displaying the compound age-specific time effect ($\beta_x\kappa_t$) for selected ages, and separately for women (left panel) and men (right panel). 
 In line with previous findings, uncertainty is larger for estimates of age-specific effect at younger ages, and particularly so among women, due to the lower frequency of death events for these populations sub-groups.


     


     

\subsection{Geographical variation in mortality by age and gender}

A key contribution of this study is that it extends the Lee-Carter model through the inclusion of a spatial effect that accounts for geographical variation in the estimates.

\begin{sidewaysfigure}
\centering

\centering

\includegraphics[width=\textwidth]{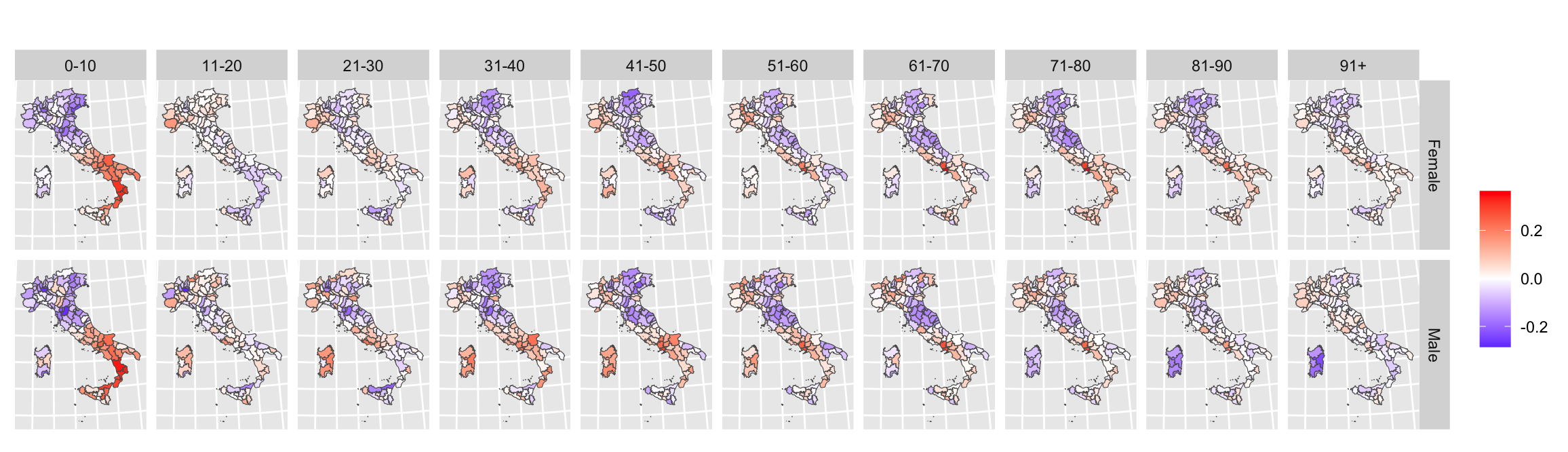}

\caption{Estimated posterior mean of the spatial effect $\omega_{sg_x}$ for each specific age class and gender. Shades of red indicate provinces with higher-than-average mortality  while shades of blue represent provinces with lower-than-average mortality.}

\label{fig:Space_Age_effect}
\end{sidewaysfigure}


 Figure \ref{fig:Space_Age_effect} maps the estimated spatial effect at the provincial level for each ten-year age class, and separately by gender. Shades of red indicate provinces with higher-than-average mortality (for each specific age class and gender); while shades of blue represent provinces with lower-than-average mortality. Figure 4bis in the Appendix \textcolor{red}{AR: I do not see the Appendix?} complements the information by providing a measure of the variance of provincial estimates. In general, it is fair to say that the darker the shades of red or blue, the more likely a province is to exhibit mortality rates which differ from the average. 
 
 At first glance, the maps highlight the existence of a gendered spatial patterning of mortality by age, characterized by two key features: first, a greater variability in male mortality rates; second, a greater variability of mortality at younger adult ages, particularly for men. The only exception to such gendered geography is observed between age zero to ten, with a marked North-South divide common to both genders, reflecting the same old tale of two Italies with very different levels of wealth, education and provision of health services.
 
 Among men, geographical differences in mortality levels are particularly pronounced between age 30 to 60, with a mortality disadvantage evident in the provinces of Sardinia and Campania and extending to other neighbouring provinces of the Centre-South (in the regions of Lazio, Calabria, Puglia and Molise). A second cluster of higher than average mortality is also found in the North-West, while provinces of the Centre-North and North-East represent areas of relatively lower mortality. Geographical disparities become less stark at older ages, although the disadvantage of Naples and other provinces of Campania, as well as that of the North-West, persists. The disadvantage of other provinces in the South fades away; Sardinia, in particular, joins the provinces of Centre-North as an area of comparatively lower mortality for men at age seventy and older.

 The geography of female mortality generally confirms the disadvantage of some provinces of the Centre-South and of the North-West, and the relative advantage of the Centre-North and North-East. Geographical disparities in female mortality rates are not as evident at younger adult ages but, contrary to what observed for men, they become more pronounced after age 60. The maps reveal the presence of a hotspot of higher than average mortality in the provinces of Naples and Caserta, and of a general disadvantage extended to all Centre-South provinces (with the exclusion of Sardinia) at older adults ages.



\subsection{Widening geographical inequalities over time}
\begin{sidewaysfigure}
\centering

\centering

\includegraphics[width=\textwidth]{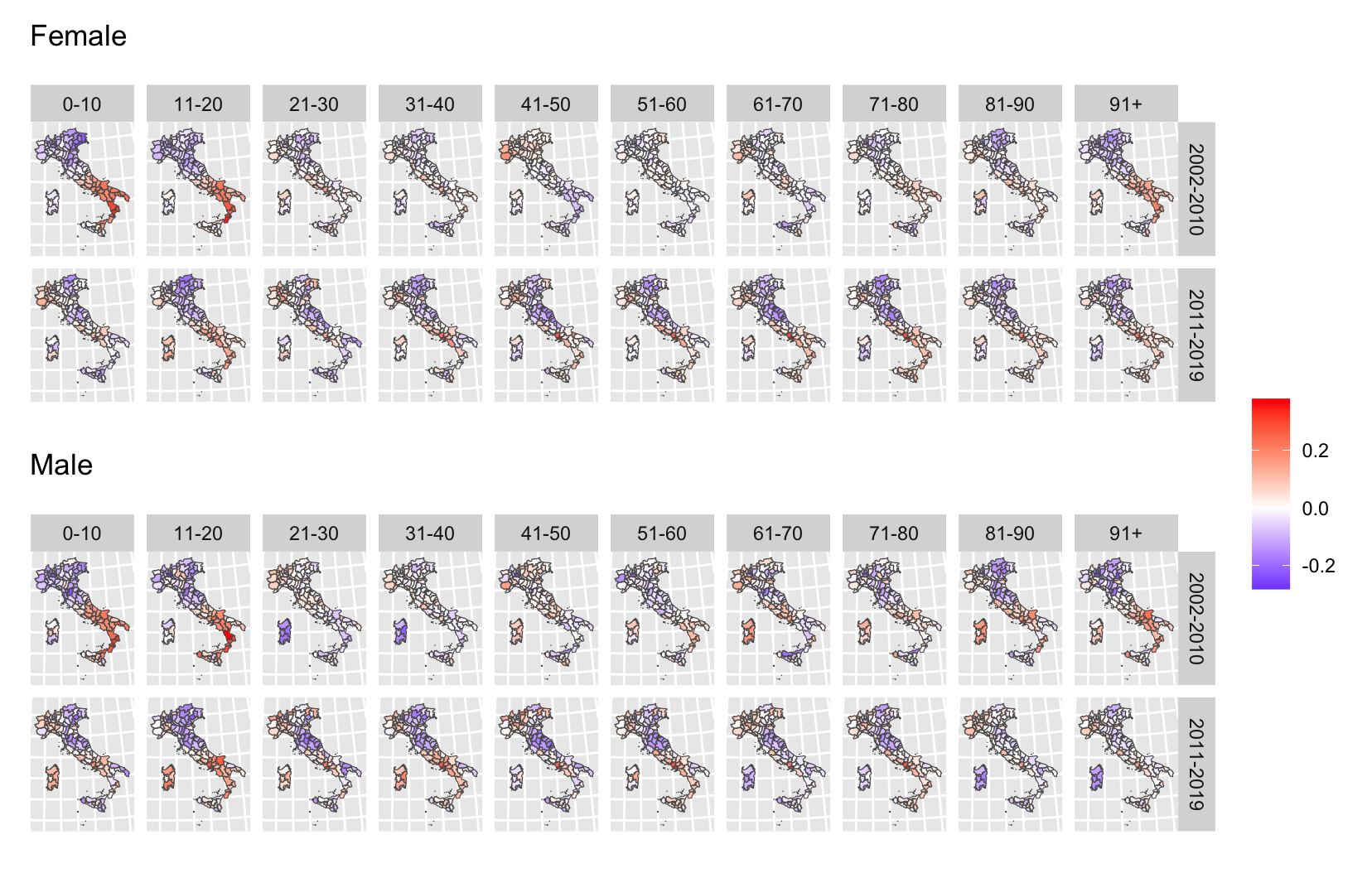}

\caption{Posterior means for the spatial effect for the second model specification for males and females. For both gender, the top panel represents the first period (2002-2010) and the bottom panel the second period (2011-2019)}

\label{fig:Space_Age_effect_model2}
\end{sidewaysfigure}

The second specification of the Model allows the spatial effect to vary across the two decades, to explicitly address the question of whether the geography of mortality by age and gender has remained constant or changed over time.
Figure \ref{fig:Space_Age_effect_model2} displays the series of age-specific spatial effects for men and  women; for both gender, the top panel represents the first period (2002-2010) and the bottom panel the second period (2011-2019). As before, provinces in blue are to be interpreted as areas of lower than average mortality, and provinces in red as areas of higher than average mortality - with darker shades signifying larger differences. Maps showing the variance of the spatial effect estimates are reported in the Appendix (Figure 5bis).

A visual comparison of age-specific maps for the two periods reveals two opposite trends, which generally hold for both genders: while geographical variability in mortality has reduced at young (up to age 20) and very old ages (after age 90), the geography of mortality at adult and older adult ages in the most recent decade has become more heterogeneous.

The marked North/South divide observed in mortality rates at younger ages over the entire period is indeed only evident in the first decade, as the disadvantage of most Southern provinces disappears in the second period, with only some moderately higher mortality in few provinces of Campania. At the same time, however, Northern provinces see a reduction of their advantage, with higher level of mortality in the North-West. A similar reduction of differences is observable for both genders, although to a lesser extent for men in their teenage years, with mortality remaining higher in the South and lower in the North.

Indeed, geographical differences in mortality rates among men widen particularly at adult ages: it is between age twenty and fifty that male mortality rates show greater variation in the second decade, with higher than average mortality in the Centre-South and North-West and a more favourable picture in the remaining Centre-North. After age fifty, the geography of mortality remains overall similar across the two periods, except for the relative improvement of the provinces of Sardinia and Marche. At older ages (80 and above), geographical differences reduce considerably.

Among women, a greater variability in the more recent period is observable mostly between age fifty and eighty. Compared to the fairly homogeneous picture of the period 2002-2010, the period 2011-2019 is characterised by pockets of higher mortality in correspondence of the provinces of Naples and Caserta in Campania (and to a lesser extent of other provinces of the Centre-South), and an extended area of comparatively lower mortality to the Centre-North-East quadrant of the map. After age 80, geographical differences are less pronounced but greater than those observed for men. It is worth noting that, in spite of this general picture of widening geographical inequalities at older ages, mortality rates for women aged 30 to 40 have also become more unevenly distributed, with higher mortality in Campania and surrounding provinces.

\section{Discussion}

This study contributes to understanding of geographical inequalities in contemporary mortality trends. It does so by leveraging recent developments in Bayesian spatial and computational statistics to move beyond aggregate data and examine trends in age- and gender-specific mortality rates at a fine geographical level. The focus is on Italy, a country with one of the highest life expectancies in the developed world, but with pronounced social and economic inequalities. These inequalities manifest geographically, shaping health and mortality patterns across a landscape far more complex than the familiar divide between the less economically developed South and the wealthier, more industrialized, and urbanized North \citep{caselli21}. As such, the country provides a salient case study to showcase the analytical potential of such methodological innovations for detailed spatio-temporal analysis.

\subsection{Recent trends in mortality by age and gender}
More precisely, by applying a spatial extension of the Lee-Carter model to estimate mortality rates across 107 Italian provinces, the study identifies, first of all, clear gender- and age-specific trends in mortality decline over the period 2002-2019. While the general level of mortality decreases for both men and women, the decline seems faster for men. Moreover, and in line with what observed across other developed countries \citep{Raleigh2019, HoHendi2018, KabirObrien, Dowd2024, Djeundje22}, our analyses lend support to the slowdown in mortality decline in more recent times, and particularly so among women. We should always avoid over-interpreting or offering simplistic explanations for what may be merely short-term fluctuations \citep{luy2020}, but we must not dismiss concerns about the slowdown as simply a sign of a ‘ceiling effect’ among higher performers \citep{Dowd2024}. Rather, if the slowdown reflects genuine shortfalls in achievable mortality improvements, analyses that account for age, gender, context (and cause) specific trends can help identify the underlying drivers and expose inequalities in these processes. 

Our estimates, indeed, suggest different paces of decline by age and gender. For example, mortality rates have overall declined faster at younger ages, particularly among men, although it should be noted that this is also the population group for which rates of mortality improvements have been much lower in the second half of the period \citep{Dowd2024}. The pace of the decline then slows down throughout older adult ages up to age 75, and more so for women. This reflects the reduced mortality improvements observed post 2010 after age 55 also by \citet{Dowd2024}, which–they argue–are largely related to stalling improvements in mortality from cardiovascular diseases. After age 75, men experience faster decline than women up to age 85, while after age 90 mortality improvements are greater for women.

\subsection{The uneven geography of mortality by age and gender}

While the findings discussed above could have been obtained with a conventional Lee-Carter model, they nevertheless provide useful context and set the stage for the subsequent two questions, which–on the contrary–could have only been addressed through its spatial counterpart, and thus constitute the unique contribution of the study: first, whether these trends manifest evenly across Italy; and second, whether the geography of mortality is becoming more heterogenous at time of slowing mortality progress.

Indeed, the estimation of the Lee-Carter model at provincial level confirms the existence of a complex mosaic, where the red and blue tiles of higher and lower mortality rates do not follow the simple North-South gradient, but appear much more scattered across the country. Moreover, their arrangement does not outline a single geography, but rather varying patterns depending on the combinations of age and gender considered \citep{Barbietal2018, caselli2003a, divino_egidi}. For example, while very few provinces in the South (i.e. Naples and Caserta) display consistently higher mortality levels across gender and age groups, the majority changes their relative positioning–such as Sardinian provinces that from areas of higher mortality at younger (male) ages become areas of lower mortality at older (male) ages; or provinces of Puglia and Sicily, which conversely turn into hotspots of higher mortality levels for women at older ages. Even the consistent relative advantage of the provinces of Tuscany, Marche and Emilia Romagna in the Centre North is not as evident at younger and older ages; and that of the Northeast shrinks among men aged fifty and older. 

Moreover, geographical variability in mortality is generally greater among younger adult men and among older adult women, suggesting the influence of age- and sex-specific, context-dependent, risk factors. These patterns likely reflect the varying importance of leading causes of death by age and gender, and how economic development, industrialization, and urbanization intersect with disparities in health care services, and differences in lifestyle and behavioural factors, to shape mortality outcomes in Italy \citep{caselli2003a}. Thus, it is not surprising that mortality among women is higher in Southern provinces at relatively older ages, when mortality from cardiovascular diseases is predominant and the role of a sound health system and effective prevention could make the difference \citep{caselli2003a, fantini2012, Barbietal2018, carboni2024,  benassi2025}; nor that mortality among men at younger adult ages is higher not just in the more disadvantaged South but also across several Northern provinces, where occupational and environmental risk factors, together with lifestyles such as alcohol consumption, may be associated with a greater relative importance of cancer mortality \citep{caselli2003a, Barbietal2018}.

A further extension of the spatial Lee-Carter model allowed the provincial effect to vary before and after 2010, revealing a widening of geographical inequalities. As a matter of fact, much of the variability discussed in the previous paragraph is driven by patterns observed in the most recent period, whereas the start of the millennium was characterised by a much more homogenous geography. This finding aligns with both earlier studies arguing that a process of convergence in regional and subregional mortality was taking place \citep{caselli2003a}; and with more recent work indicating, on the contrary, the end of such convergence and the widening of spatial inequalities in the second decade of the 2000s, in correspondence with slowing survival improvements \citep{Hrzic2020, carboni2024, Sauerberg2024, camarda25}.

In summary, this study extends a commonly applied model to the analysis of temporal trends in mortality rates \citep{lee-carter92, BASELLINI2023} to account for the existence of a geographical structure to such trends. It overcomes limitations imposed by random fluctuations of death counts across small areas such as the Italian Provinces by adopting a Bayesian approach that borrows strength from neighbouring observations to reduce uncertainty of the estimates; and it demonstrates the value of recent computational advancements by employing {\tt inlabru}, an {\tt R} software package that uses Integrated Nested Laplace approximation \citep{rue2009approximate} for the estimation of complex spatial models with non-linear effects. Through this approach, it contributes to empirical literature on contemporary mortality trends by highlighting the age patterning of the gendered geography of mortality in Italy, as well as the widening of spatial inequalities in more recent times.

\subsection{Limitations and future research avenues}

Notwithstanding the important contribution of the present study, there are some limitations that warrant consideration and indicate opportunities for future developments.

While our results suggest that different factors may be conditioning mortality of men and women in the different areas of the country, our analysis did not explicitly assess the association of mortality rates with socio-economic or environmental factors. Existing studies on selected time periods demonstrated, consistently at different geographical scales, the existence of significant associations between mortality and various measures of socio-economic vulnerability, such as for example: unemployment rates at municipal level \citep{santossanchez2020}, a composite index of socio-economic vulnerability  at provincial level \citep{Barbietal2018}, absolute and relative income at regional level \citep{dallolio2012}. The spatial Lee-Carter model could indeed have been extended to include geo-referenced covariates. However, identifying relevant and insightful variables collected annually, in a consistent format over an extended period, and at the provincial level, is not straightforward, and would have suffered from comparability issues due to the changing provincial boundaries over time.

More interestingly, operating within a Bayesian framework would have allowed us to produce accurate and reliable estimates at even finer levels of detail as, for example, disaggregating mortality rates also by main (or grouped) causes of death. Unfortunately, the risk of disclosing personal information meant that the Italian National Institute of Statistics could not grant us access to such data. We would welcome future collaborations with National Statistical Institutes on the geographical analysis of age- and cause-specific mortality. Such granular estimates would contribute to uncover the social, structural and environmental nature of place-specific drivers of mortality \citep{caselli2003a, Barbietal2018}, providing invaluable information to tailor place-based public health policies and interventions to the populations most at risk.

This study estimated the spatial Lee-Carter model separately by gender, as it is common praxis. In principle, it would have been possible to break down the population by other socio-economic characteristics (such as educational level)–thus allowing an exploration of social as well as geographical inequalities. While often information on individual characteristics is not routinely collected or accurately recorded in vital events data, future applications of the method could leverage the opportunity provided by their linkage to census records, or to other administrative database. These data sources offer key advantages for studying health and mortality inequalities, including large sample sizes that enhance precision—especially for rare conditions or minority groups—and long timeframes well suited to tracking changes over time \citep{keenanetal_21}. Mortality rates calculated on linked/register data also rely on the same source for status variables in both their numerator and denominator, thus producing more reliable estimates. 

Several countries in Europe routinely link their vital events data to census or population register—see, for example, \citet{keenanetal_21} for a showcase of studies on spatial and social inequalities in health and mortality using register data from Sweden, Finland, Belgium, Lithuania and from the four constituent countries of the United Kingdom. The Italian National Institute of Statistics does not, to our knowledge, systematically link vital events to Census or Population Registers but has often performed ad-hoc linkages contributing valuable insights to the study of mortality inequalities. For example, the linkage of mortality data for the years 2012-2014 to the 2011 Population Census \citep{Murtin2017, istat16} showed that low education is, as expected, a key driver of premature mortality but with differences in severity by gender, geography and cause of death. While educational inequalities in alcohol and tobacco related deaths are particularly pronounced among men, diabetes is associated to larger inequalities among women, particularly in the South \citep{istat17}. 
Given the value of these data linkages, along with advances in computational and statistical methods that address challenges in complex spatio-temporal models and sparse data, we strongly support the Institute’s continued efforts\footnote{for example, we are aware of the Mortality Inequality Database 2019–2020, linking deaths from the National Register of Causes of Death to the National Base Register; and of ongoing work integrating the Istat Population Register with the Income Statistical Register.} to integrate vital records and population registers for tracking social and spatial inequalities in mortality.

\section{Conclusion}

Our study demonstrated the value of such granular approaches, highlighting the widening of geographical inequalities in mortality rates after 2010, with a distinct age and gender patterning, being more pronounced at younger adult ages for men and older adults ages for women. Recent and future trends may be further compounded by the mortality shock and the long-lasting health consequences of the Covid-19 pandemic. The future of mortality in high-income countries is–at best–uncertain \citep{Dowd2024}. To understand current trends and anticipate (and plan for) future ones, it is thus of even greater importance to move beyond aggregate data and analysis to understand how mortality varies between population groups and across regions, and to gain insights on the underlying factors responsible for such variation that may be holding back progress for some.

\bibliography{sn-bibliography}

\end{document}